# Home Location Identification of Twitter Users[1]

JALAL MAHMUD, IBM Research – Almaden
JEFFREY NICHOLS, IBM Research – Almaden
CLEMENS DREWS, IBM Research – Almaden

We present a new algorithm for inferring the home location of Twitter users at different granularities, including city, state, time zone or geographic region, using the content of users' tweets and their tweeting behavior. Unlike existing approaches, our algorithm uses an ensemble of statistical and heuristic classifiers to predict locations and makes use of a geographic gazetteer dictionary to identify place-name entities. We find that a hierarchical classification approach, where time zone, state or geographic region is predicted first and city is predicted next, can improve prediction accuracy. We have also analyzed movement variations of Twitter users, built a classifier to predict whether a user was travelling in a certain period of time and use that to further improve the location detection accuracy. Experimental evidence suggests that our algorithm works well in practice and outperforms the best existing algorithms for predicting the home location of Twitter users.

ACM Classification: **H5.2 [Information interfaces and presentation]**: User Interfaces. **H2.8 [Database Management]**: Database Applications – Data Mining.

General Terms: Algorithms, Design, Experimentation, Human Factors

Additional Key Words and Phrases: Location, Tweets, Time zone.

## 1. INTRODUCTION

Recent years have seen a rapid growth in micro-blogging[2] and the rise of popular micro-blogging services such as Twitter. As of March 21, 2013, 400 million tweets were being posted everyday[3]. This has spurred numerous research efforts to mine this data for various applications, such as event detection [Sakaki et al. 2010; Agarwal et al. 2012], epidemic dispersion [Lampos et al. 2010], and news recommendation [Phelan et al. 2009]. Many such applications could benefit from information about the location of users, but unfortunately location information is currently very sparse. Less than 1% of tweets are geo-tagged[4] and information available from the location fields in users' profiles is unreliable at best. Cheng et al. found that only 26% of Twitter users in a random sample of over 1 million users reported their city-level location in their profiles and only 0.42% of the tweets in their dataset were geo-tagged [Cheng et al. 2010]. Hecht et al. report that only 42% of Twitter users in their dataset reported valid city-level locations in their profile and 0.77% of the tweets were geo-tagged [Hecht et al 2011].

In this paper, we aim to overcome this location sparseness problem by developing algorithms to predict the *home*, or primary, locations of Twitter users from the content of their tweets and their tweeting behavior. Ultimately, we would like to be able to predict the location of each tweet and our work to predict a user's home location is a key step towards achieving that goal. This is because single tweets rarely contain enough information by themselves to reliably infer a location. Knowing the user's home location gives an important clue to the possible location of a tweet, and we expect in the future that this information will be combined with other inferred information, such as the likelihood that user is traveling (which we also explore briefly here), to infer a location for a single tweet.

---

[1] This paper is an extended version of [Mahmud et al. 2012]
[2] http://www.businessinsider.com/nobody-blogs-anymore-theyres-all-microblogging-2011-2
[3] http://articles.washingtonpost.com/2013-03-21/business/37889387_1_tweets-jack-dorsey-twitter
[4] http://thenextweb.com/2010/01/15/Twitter-geofail-023-tweets-geotagged/

Our goal is to predict home location at the city-level, though we also examine the possibility of predicting at other larger levels of granularity, such as state, time zone and geographic region. The benefit of developing these algorithms is two-fold. First, the output can be used to create location-based visualizations and applications on top of Twitter. For example, a journalist tracking an event on Twitter may want to know which tweets are coming from users who are likely to be in a location of that event, vs. tweets coming from users who are likely to be far away. As another example, a retailer or a consumer products vendor may track trending opinions about their products and services and analyze differences across geographies. Second, our examination of the discriminative features used by our algorithms suggests strategies for users to employ if they wish to micro-blog publicly but not inadvertently reveal their location.

Our research is motivated by a variety of previous work on home location inference from tweets [Eisenstein et al. 2011; Hecht et al. 2011; Cheng et al. 2010; Chang et al. 2012; Chandar et al. 2011; Kinsela et al. 2011]. A few also attempt to predict the home location of users at the city-level [Cheng et al. 2010; Chang et al. 2012; Chandar et al. 2011; Kinsela et al. 2011]. City-level location detection is more challenging than detecting location at a higher granularity such as state or country. Cheng et al. [2010], Chang et al. [2012], and Chandar et al. [2011] reported city-location detection accuracy using an approximate metric, where a prediction is deemed as correct if it is within 100 miles of the actual city-location. Using such a relaxed accuracy metric, the best city-location detection accuracy is reported as approximately 50% [Cheng et al. 2010; Chang et al. 2012]. On the other hand, Kinsela et al. [2011] reported 32% (exact) accuracy for city-location detection. We improve on these results in our work.

In particular, we make the following contributions:

- An algorithm for predicting the home location of Twitter users from tweet contents, tweeting behavior (volume of tweets per time unit), and external location knowledge (e.g., dictionary containing names of cities and states, and location based services such as Foursquare[5]). Our algorithm leverages explicit references of locations in tweets (such as mentions of cities or states within the tweets), but still works with reduced accuracy when no such explicit references are available. Our algorithm uses an ensemble of several classifiers.

- An algorithm for predicting locations hierarchically using time zone, state or geographic region as the first level and city at the second level. Our approach has the promise to be used as an infrastructure for more granular location predictions in the future.

- An evaluation demonstrating that our algorithm outperforms the best existing algorithms for home location prediction from tweets. Our best method achieves accuracies of 64% for cities, 66% for states, 78% for time zones and 71% for regions when trained and tested using a dataset consisting of 1.52 million tweets from 9551 users from the top 100 US cities. We also demonstrate using the dataset of Cheng et al. [2010] that our best method outperforms their method for users' city-level home location prediction.

---

[5] http://www.foursquare.com/

- An analysis of movement variations of Twitter users and correlation with location prediction, which confirms our hypothesis that locations are less accurately predictable for frequently traveling users. Based on the analysis, we present an algorithm for detecting travelling users and use the result of the algorithm to improve the location prediction accuracy. When users identified as travelling are eliminated, location prediction accuracy improves to 68% for cities, 70% for states, 80% for time-zones and 73% for regions.

In the remainder of this paper, we discuss related work, our data set, a formalization of the location estimation problem, our location classification approaches, ensemble approaches, and an evaluation of our algorithms. Then, we present our analysis of movement and location prediction. Finally, we conclude the paper with a discussion of future research.

## 2. RELATED WORK

Our research is related to a variety of prior work in the following areas:

### 2.1 Content-based Location Estimation from Tweets

A number of algorithms have been proposed to estimate the home location of twitter users using content analysis of tweets [Eisenstein et al. 2011; Hecht et al. 2011; Cheng et al. 2010; Chang et al. 2012; Chandar et al. 2011; Kinsela et al. 2011]. One commonality among those methods is that they build probabilistic models from tweet content.

Eisenstein et al. [2011] built geographic topic models to predict the location of Twitter users in terms of regions and states. They reported 58% accuracy for predicting regions (4 regions) and 24% accuracy for predicting states (48 continental US states and the District of Columbia). Hecht et al. [2011] built Bayesian probabilistic models from words in tweets for estimating the country and state-level location of Twitter users. They used location information submitted by users in their Twitter profiles, resolved via the Google geolocation API, to form the ground-truth of a statistical model for location estimation. They were able to get approximately 89% accuracy for predicting countries (4 countries), but only 27% accuracy for predicting states (50 states in US). The higher accuracy reported for predicting country was largely due to the uneven distribution of countries in their dataset, where 82% users were from US and hence a US-only predictor could also achieve 82% accuracy for predicting countries using that dataset.

City-level location estimation is more challenging than location estimation at higher granularities, such as states or countries, because the number of cities in a typical dataset is often much larger than the number of states, regions, or countries. City-level home location estimation is described in [Cheng et al. 2010; Chang et al. 2012; Chandar et al. 2011; Kinsela et al. 2011].

Cheng et al. [2010] describe a city-level location estimation algorithm, which is based on identifying *local words* (such as "red sox" is local to "Boston") from tweets and building statistical predictive models from them. However, their method requires a manual selection of such local words for training a supervised classification model. While they report approximately 51% accuracy using their approach, their accuracy metric is relaxed such that the actual city could be anywhere within 100 miles from the predicted city. When an exact city-level prediction was required, accuracy dropped to less than 5%. Chandar et al. [2011] described location estimation using the conversation relationship of Twitter users in addition to the text content used in the conversation. They have used a subset of the dataset collected by Cheng et al. [2010], and reported 22% accuracy in correctly predicting city-level locations within 100 miles of actual city-location.

More recently Chang et al. [2012] described yet another content based location detection method using Gaussian Mixture Model (GMM) and the Maximum Likelihood Estimation (MLE). Their method also eliminates noisy data from tweet content using the notion of non-localness and geometric-localness. Their approach selected local words using an unsupervised approach and achieved approximately 50% accuracy in predicting city-location within 100 miles of actual city-location, which is comparable to Cheng et al. [2010].

Cheng et al. [2010], Chandar et al. [2011] and Chang et al. [2012] reported city-location detection accuracy using an approximate metric (e.g., accuracy within 100 miles). However, Kinsela et al. [2011] reported location detection at various granularities using an exact accuracy metric (whether the detected location matches the actual location) using a language modeling approach to build models of locations. Their algorithm can predict the location of a tweet (from which location the tweet originated) as well as the location of a Twitter user, when her tweets are aggregated for a given period. For building the language models of locations, they used geo-tagged tweets originating from those locations. For predicting tweet-level location, they achieved 53% accuracy for country level, 31% accuracy at state level, 30% accuracy at city level and 14% accuracy at zip-code level. For user-level location prediction, accuracies are 76% for country, 45% for state, 32% for city and 15% for zip-code. In this work, we do not predict location at the zip-code level, however our accuracies for predicting location at higher granularities are better than the accuracies reported by Kinsela et al. [2011].

Our work makes use of some these findings while also going further. In particular, our content-based statistical classifier also uses a Bayesian model of local word distributions to predict location, similar to Hecht et al. [2010] and Cheng et al. [2011]. This classifier is just one of several that we use in our ensemble however, and we have improved classification further by using a hierarchy of classifiers that predict location at different granularities. In terms of accuracy, we experimentally demonstrate that our algorithm achieves higher accuracy for detecting states, regions and cities than existing algorithms. Previous work did not consider classifying the time-zone of a user, which we have added in our work. Our behavior based time-zone classifier uses novel temporal behavior-based features not used by any existing work. We have also experimentally compared the performance of our algorithm for city-level location prediction with that of Cheng et al [2010] using their dataset.

**2.2 Content-based Location Extraction from Tweets**

Content-based methods have also been used to determine the geo-location of a tweet or to extract location information from tweets.

Dalvi et al. [2012] studied the problem of matching a tweet to an object, where the object is from a list of objects in a given domain (e.g., restaurants). They assume that the geo-location of such objects is already known. Their model utilizes such geographic information using the assumption that the probability of a user tweeting about an object depends on the distance between the user's location and the object's location. Such matching can also geo-locate tweets and infer the present location of a user based on the tweets about geo-located objects. Along the same line, Li et al. [2011] describe a method to associate a single tweet to points of interests, such as a restaurant, shop or park, by building a language model for each point of interest and using standard techniques such as KL-Divergence. This is dependent on availability of enough tweets for each point of interest, and is different from estimating the home location of a user.

Recently Agarwal et al. [2012] describe a dictionary-based method for extracting location information from tweets. They use named entity recognition as well as a concept vocabulary-based method to identify words that denote a location name from tweets. For disambiguating place names, they use a machine learning method, an inverted index search on World Gazetteer data, and search using the Google Maps API. Extracting location-names from tweets is a first step to building our place-name based classifier, and disambiguation methods for extracting place names as described in Agarwal et al. [2012] is complementary to our work.

Our goal is to estimate home location of a Twitter user. Location extraction from tweets is a different problem from estimating the home location of Twitter users. However, some of the methods for matching tweets to a geo-location can be used for feature extraction for home location estimation from tweets.

**2.3 Location Estimation without using Tweets Content**

There are efforts to estimate the location of Twitter users using location information provided in Twitter profile, geo-tagged tweets and social network information.

A number of works make use of location information submitted by users in their Twitter profiles. For example, Kulshrestha et al. [2012] have used location information reported in twitter user's profile and multiple map APIs to find location of users at country level for further analysis. They compared location information provided by multiple map APIs to reduce inference errors. In this way, they were able to infer country-level location of 23.5% of users with 94.7% accuracy. However, these techniques of location inference rely on the users themselves, whereas a large number of such users either enter incorrect non-geographic information in the location field of their profile or leave the field empty (34% as reported by Hecht et al. [2011]). In addition, map APIs do not always return the correct result.

Recently, Sadilek et al. [2012] described a location estimation method that can infer the most likely location of people for a given time period from the geo-location information of their friends for that time period. The assumption is that location information of friends is shared through GPS-enabled devices or location-based services, such as Foursquare. They have implemented both a supervised and unsupervised version of their algorithm. In their supervised approach, previously visited locations of users are also given to the prediction algorithm in addition to their friends' locations. In the unsupervised approach, such information (user's previous visited locations) is not given to the algorithm. For the unsupervised approach, they have demonstrated that when a person has at least 2 geo-active friends for whom geo-information of tweets are available, the location of the person can be predicted at the neighborhood level (e.g., a foursquare venue) with 47% accuracy using their algorithm and when 9 geo-active friends' information is available, location can be predicted with 57% accuracy. These accuracies are higher with supervised approach (77% with 2 friend's information and 84% for 9 friend's information). However their approach is dependent on one's geo-active friends (who post messages with geo-location at-least 100 times a month), and the availability of geolocation information for such friends for a given period. In addition, their location prediction algorithm also assumes that a set of locations (e.g., foursquare venues) are frequently visited by users. These assumptions may not be valid for many users who do not have such friends or do not frequently visit such popular locations.

We believe that location detection using Twitter content, Twitter profile information, geo-active users' information and social network are complementary efforts. Such approaches may be combined together to further increase accuracy, perhaps using the ensemble approach we introduce here.

### 2.4 Location Estimation from other Social Media

Lieberman and Lin [2009] used geo-pages in Wikipedia to infer locations (as granular as small geographic regions) of their contributors. Popescu and Grefenstette predict the home country of Flickr users by analyzing manual annotations with place names and geo-tags [Popescu et al. 2010]. In contrast, our location inference algorithm does not use manual annotations or geo-tags, although geo-tags might be employed in the future to improve the accuracy of the algorithm. Backstrom et al. [2010] used the social network structure of Facebook to predict location of Facebook users. We do not currently use social network features in our algorithm, but these could be incorporated in the future.

Recently Chang et al. [2011] describe a system that can predict places where users will go next based on their previous check-ins with the Facebook Places service. Along the similar line, Gao et al. [2012] has explored the pattern of user check-ins on location-based social networks, such as Foursquare, and built a predictive model for users' check-in behaviors. Their main finding is that users with friendship tend to go to similar locations than those without, and users' visits follow a power-law distribution, which means they tend to visit few places many times and many other places few times. Cho et al. [2011] describes modeling user location from the location-based social network Gowalla[6]. They have developed a periodic and social mobility model for predicting the mobility of users (e.g., when user is at "home" and when user is at "work"). Their result suggests that there is a strong periodic behavior throughout certain periods of the day alternating between primary (e.g., "home") and secondary (e.g., "work") locations on weekdays, and "home" and social network driven locations on weekends. Their work uses the check-in history of users' and their friends from a location-based social network. Their focus is not to detect the home location of Twitter users, but rather detecting their mobility patterns.

It is uncommon for us to have access to exact location information for a user, although our algorithm can use it if available (e.g., a tweet generated by Foursquare that notes the user's exact location). Instead our algorithm must rely on a variety of features collected from users' recent tweets. Furthermore, our focus is to predict users' home locations instead of their potential future locations.

---

[6] http://www.gowalla.com

Cranshaw et al. [2012] has described a clustering based model to understand the structure and composition of a city based on the social media its residents generate through location based services such as Foursquare. Their method can discover distinctly characterized areas of a city (such as neighborhoods) by using the spatial proximity of venues users check-in to and the social proximity of users. A number of researchers have used named entity detection with a geographical gazetteer for location estimation from blog posts [Fink et al. 2009] and web pages [Amitey et al. 2004; Zong et al. 2005]. We also identify names of cities and states from tweets using the USGS gazetteer and use them to build statistical models. As we will discuss later, we have found that using these terms alone does not give the best accuracy. Adams et al. [2012] has described a method to estimate geographic regions from unstructured, non geo-referenced text by combining natural language processing, geo-statistics, and a data-driven bottom-up semantics. They use the hypothesis that natural language expressions are geo-indicative even without explicit reference to place names (such as manufacturing, traffic, employment is more indicative of a larger city). They have applied their algorithm on large text documents such as blog and Wikipedia article where there are few coherent topics. This approach may not work for tweets, which are noisy and do not have a coherent topics.

### 3. DATASET

From July 2011 to Aug 2011, we collected tweets from the top 100 cities in US by population[7] (see Figure 1). First, we obtained a bounding box in terms of latitude and longitude for each city using Google's geo-coding API[8]. We recorded tweets using the geo-tag filter option of Twitter's streaming API[9] for each of those bounding boxes until we received tweets from 100 unique users in each location. The city corresponding to the bounding box where the user was discovered was assumed to be the ground truth home location for that user. We discuss the validity of this assumption later in the paper.

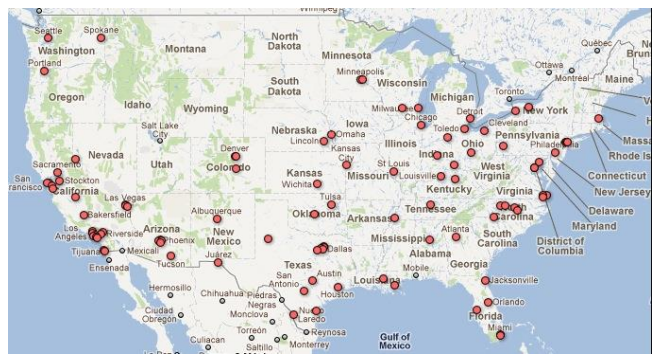

**Figure 1. Cities in our dataset**

We then invoked the Twitter REST API[10] to collect each user's 200 most recent tweets (less if that user had fewer than 200 total tweets). Some users were discovered to have private profiles and we eliminated them from our dataset. Our final data set contains 1,524,522 tweets generated by 9551 users[11]. 100599 tweets (6.6%) were generated by Foursquare and contained URLs that could be accessed to retrieve exact location descriptions. 289650 tweets (19%) contained references to cities or states mentioned in the USGS gazetteer[12], however this number also includes ambiguous matches (e.g., the word "black" being matched as a town in Alabama) and the Foursquare tweets that also often contain textual references to cities or states. We divided the entire dataset into training (90%) and testing (10%) sets for 10-fold cross-validation.

---
[7] http://en.wikipedia.org/wiki/List_of_regions_of_the_United_States
[8] http://code.google.com/apis/maps/documentation/geocoding/
[9] http://dev.twitter.com/pages/streaming_api
[10] http://en.wikipedia.org/wiki/List_of_United_States_cities_by_population
[11] This data set is available through the ICWSM Dataset Sharing Service at http://icwsm.cs.mcgill.ca/

|    | Tweets |
|----|--------|
| 1. | Foldin the biggest pile of clothes EVER!! |
| 2. | Let's cruise on today (@ House of Ambrose) http://4sq.com/m1F3R5 |
| 3. | #Portland It's chocolate peanut butter! |
| 4. | Let's Go Red Sox!!! |
| 5. | Another sunny day in California! |

**Table 1. Example tweets in our dataset**

## 4. LOCATION ESTIMATION – PROBLEM STATEMENT

For this paper, we denote location of a user u at granularity g as $L_g(u)$, where

$$L_g(u) = f(S_u, T_u, E)$$

Thus $L_g(u)$ is a function of $S_u$, $T_u$ and E. $S_u$ represents the set of tweets for that user, $T_u$ represents the set of creation times of those tweets, and E represents the set of external location-based knowledge available from a location based service, such as Foursquare, or a dictionary, such as the USGS gazetteer. In this equation, the desired granularity can be at any level, including country, state, geographic region, time zone, city, street or land mark. The parameter E is optional since external knowledge may not always be available. For all granularities except time zone, $S_u$ is mandatory. Time zone estimation is possible with only $T_u$, however $S_u$ and E can also be used for estimation if they are available. In this paper, without loss of generality we present location estimation for the following granularities: city, time zone, state and geographic region.

## 5. LOCATION CLASSIFICATION APPROACHES

Here we describe each of our location classifiers in detail.

### 5.1 Content-based Statistical Classifiers

We use three statistical location classifiers that are each trained from different terms extracted from S, the set of all users' tweets. The classifiers and their associated terms are:

- **Words:** all words contained within S
- **Hashtags:** all hashtags contained within S
- **Place Names:** all city and state location names within S, as identified via a geographical gazetteer

---

[12] http://www.census.gov/geo/www/gazetteer/places2k.html.

These classifiers can be created for any level of location granularity for which we have ground truth. Each user in our training dataset corresponds to a training example, where features are derived from his or her tweet contents. The output is a trained model with the number of classes equal to the total number of locations of that granularity in our training dataset (e.g., total number of cities). All of these classifiers use the same approaches for feature selection, training, and classification, which are described below.

| Term | Location Type | Location Conditional Distribution | Local? |
| --- | --- | --- | --- |
| Grass | City | Houston:0.31<br>Boston:0.23<br>Fresno:0.16<br>Tulsa:0.15<br>Pittsburgh:0.15 | No |
| Vegas | Time Zone | Pacific: 0.8588<br>Eastern: 0.0705<br>Mountain:0.0470<br>Central: 0.0235 | Yes |

**Table 2. Examples of local and non-local terms with their conditional distributions**

*5.1.1 Feature Extraction*

First, we tokenize all tweets in the training dataset, which removes punctuation and other whitespace. All URLs and most tokens containing special characters are then removed, except for tokens that represent hashtags and start with # (e.g., the token #Portland in Table 1).

Once the tokens have been extracted, different processes are used to extract terms for each classifier. For the Words classifier, we use as terms all tokens that are not identified as stop words or marked as nouns by a part-of-speech tagger. Stop words are defined by a standard list of 319 stop words, and parts of speech are classified using Open NLP[13]. We do not use adjectives, verbs, prepositions, etc. because they are often generic and may not discriminate among locations. For the Hashtags classifier, we use as terms all tokens that start with the # symbol. For the Place Names classifier, we generate a set of terms that appear in the tweets and match names of US cities and states from the USGS gazetteer. Not all city or state names are a single word, so we first generate bi- and tri-grams from the ordered list of tokens. We then compare all uni-, bi-, and tri-grams to the list of city and state names. Any matching names are used as terms. Note that some common words are used as the names of cities in the US (e.g., "eagle" is a town in Colorado, and "black" is a town in Alabama). We do not currently attempt to distinguish between uses of common words to refer to locations compared to their usual meanings.

Once we have the set of terms for a particular classifier, it is helpful to identify terms that are particularly discriminative (or "local") for a location (also discussed by Cheng et al. [2010]). For example, we found that the term "Red Sox", extracted from the 4th tweet in Table 1, is local to the city "Boston." We use several heuristics to select *local terms*. First, we compute the frequency of the selected terms for each location and the number of people in that location who have used them in their tweets. We keep the terms that are present in the tweets of at least K% people in that location, where K is an empirically selected parameter. We experimented with different values and selected K=5. This process also eliminates possible noisy terms. Next, we compute the average and maximum conditional probabilities of locations for each term, and test if the difference between these probabilities is above a threshold, $T_{diff}$. If this test is successful, we then further test if the maximum conditional probability is above a threshold, $T_{max}$. This ensures that the term has high bias towards a particular location. Applying these heuristics gives us localized terms and eliminates many terms with uniform distribution across all locations. We set these thresholds empirically at $T_{diff} = 0.1$ and $T_{max} = 0.5$. Table 2 shows a few terms and their conditional distributions. These local terms become features for our statistical models.

---

[13] http://opennlp.sourceforge.net/projects.html

*5.1.2 Training and Classification*

Once the features (i.e. local terms from the previous step) are extracted for each classifier, we build statistical models using standard machine learning approaches. We have tried a number of classifiers from WEKA[14] such as Naïve Bayes, Naïve Bayes Multimonial, SMO (an SVM implementation), J48, PART and Random Forest. We found that Naïve Bayes Multimonial, SMO and J48 classifiers produced reasonable classification results for our dataset; we empirically selected Naïve Bayes Multimonial.

## 5.2 Content-based Heuristic Classifiers

We have also built two heuristic classifiers that predict users' locations at different granularities. The *local-place* heuristic classifier is specific to classifying city or state-level location. The heuristic is that a user would mention his or her home city or state in tweets more often than any other cities or states. For every city or state in our training corpus, we compute the frequency of its occurrences in user's tweets and use this as the matching score of that user with that city or state. The city or state with the highest matching score is predicted as the location classification for that user.

The *visit-history* heuristic classifier is applicable to location classification at all granularities. The heuristic is that a user would visit places in his home location more often than places in other locations. In order to retrieve a user's visit history, we look for URLs generated by the Foursquare location check-in service in their tweets (e.g., the 2nd tweet in Table 1 contains one such URL), retrieve venue location information from those URLs (e.g., city, state) using the Foursquare API, and build a frequency-based statistic for the visited locations at the desired level of granularity. Foursquare venues typically contain detailed low-level location information, so a location value at the correct level of granularity can usually be determined. Links that cannot be resolved to a venue are discarded. The location with the highest frequency is returned as the location classification for the user.

## 5.3 Behavior-based Time Zone Classifier

We hypothesize that users tweeting behavior follow certain patterns. For example, certain periods of the day may have more tweeting activity than others. However, such behavior is also dependent on their time zones when we consider a specific time

---

[14] http://www.cs.waikato.ac.nz/ml/weka/

(e.g., GMT+6). As an example, consider 8pm Eastern time. A user who lives in NY in the Eastern time zone is more likely to be tweeting (since s/he may already have returned from work) at that time in comparison to a user who lives in CA (for whom it is 5pm and s/he may be still be at work). When we compare tweet creation times for users in different time zones, we hope to discover temporal shifts in tweeting activity. Figure 2(a) shows the average tweet volume per user for each hour of the day in the four US time zones. All tweet creation times are recorded and shown in GMT. From this graph, it can be seen that tweet behavior throughout the day has the same shape in each time zone and that there is a noticeable temporal offset that a classifier should be able to leverage to predict the time zone for a user. To construct the classifier, we first divide the day into equal-sized time slots of a pre-specified duration. Each time slot represents a feature-dimension for the classifier. We have tried different sizes for time slots, e.g., 60, 30, 15, 5, and 1 minutes. We empirically chose 1 minute duration time slots for our classifier. For each time slot, we count the number of tweets sent during that time slot for each user in our training set. Since total tweet frequency in a day varies across users, we normalize the number of tweets in a time slot for a user by the total number of tweets for that user.

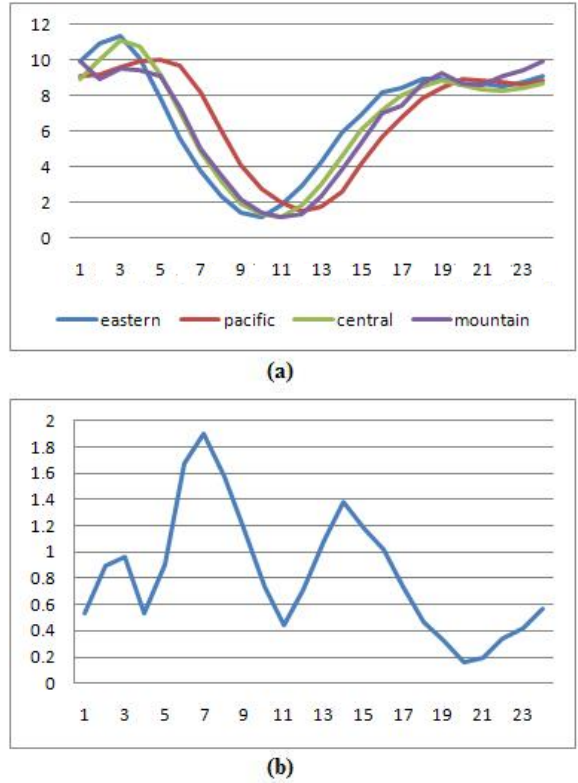

**Figure 2. (a) Variations of average tweet volume/user across time zones for different hours of the day (b) Variations of standard deviations of average tweet volumes across time zones for different hours of day**

Figure 2(a) shows that the differences between tweet volumes in different time zones are not uniform throughout the day. For example, there is little difference in tweet volume across all of the time zones at hour 11. On the other hand, there is large difference in tweet volume across all of the time zones at hour 3 or hour 7. Figure 2(b) shows variations of standard deviations of tweet volumes across time zones. These variations show that different times of the day are more discriminative. For example, hour 7 seems to have high variation of standard deviations of average tweet volumes from different time zones. Thus, this time slot is quite discriminative to differentiate time zones by their average tweet volumes. We capture this variation in our model by weighting the feature values of each time-slot using the standard deviation for that time slot. To train the classifier, we use the Naïve Bayes classifier from WEKA.

## 6. ENSEMBLE OF LOCATION CLASSIFIERS

We also create an ensemble of our classifiers to improve accuracy. In machine learning, multiple classifiers are often combined in an ensemble [Dietterich et al. 2010; Rokach et al. 2010], which is often more accurate than creating an individual classifier in the ensemble. Among the ensemble methods, majority voting is the simplest where the final classification is the class that receives the most votes from individual classifiers [Rokach et al. 2010]. Bagging is another approach, which is based on re-sampling the training dataset to learn individual classifiers and then using majority vote to combine classifications [Breiman et al. 1996]. More complex approaches are also used, such as boosting, where each classifier receives a weight that is learned based on classifier performance [Freund et al. 1996]. Note in this approach that the weights are learned once and then remain static when the trained classifier is used to classify new instances. There is also the dynamically weighted ensemble method, which aggregates the outputs of multiple classifiers using a weighted combination where each weight is based on the certainty of the respective classifier for classifying that instance [Jiménez et al. 1998]. Using this method, the classifier weights change dynamically based on the properties of the instance being classified.

In this work, we use a dynamically weighted ensemble method to create an ensemble of the statistical and heuristic classifiers for two reasons. First, we wanted to choose an ensemble method that would account for the differences in the available information for each user. For example, some users may have many Foursquare check-in tweets, which favors the visit history classifier, whereas others may use many location words, favoring the local term and place name classifiers. The dynamically weighted ensemble method weights each classifier differently for each instance based on a confidence estimate or certainty of that classifier for classifying that instance (different from fixed or static weighting) [Jiménez et al. 1998]. Second, we wanted to choose an ensemble method that would allow us to include both the statistical and heuristic classifiers. Several widely known ensemble methods, such as bagging and boosting, require multiple iterations of re-sampling and re-training the component classifiers, which is only possible for statistical classifiers.

Here we will introduce a metric, *Classification Strength,* which we use in our dynamically weighted ensemble implementation. Let T denote the set of terms from user's tweets that would be considered for classification using a particular classifier. For statistical classifiers, the *matching location set* is the set of locations in our trained model containing terms from T. For the local-place classifier, this set contains locations from our dataset that match content in the user's tweets. For the visit-history classifier, this set contains locations from the user's visit history that appear in our dataset. The *Classification Strength* for a user is the inverse of the number of possible matching locations in the matching location set. Thus, if more locations are contained in the matching location set, the classification strength will be lower and vice versa. As a concrete example, suppose that a user's tweets contained words that match to five different cities, thus producing the matching location set: {New York, Los Angeles, Chicago, Dallas, Boston}. The classification strength for this set is 1/5 = 0.2.

The classification strength of a classifier for a particular instance expresses the discriminative ability of that classifier for classifying that instance. For our implementation, the classification strength of a classifier for a particular instance is used as the weight of that classifier in the ensemble for classifying that instance. For the behavior-based time zone classifier, we use the confidence value of the classification for a particular instance as its weight.

In order to validate that our dynamically weighted ensemble approach is correct, we also created ensembles using two other techniques: majority voting [Rokach et al. 2010] and multi-class AdaBoost [Zhu et al. 2009], which is an extension from the original AdaBoost algorithm [Freund et al. 1996]. As discussed previously, it was not possible to include the heuristic classifiers in the ensemble using AdaBoost, so this ensemble uses only the statistical classifiers. For completeness, we constructed two majority voting ensembles, one using only the statistical classifiers and another using both the statistical and heuristic classifiers. A comparison of the results of these different methods is presented later in the Experiments section.

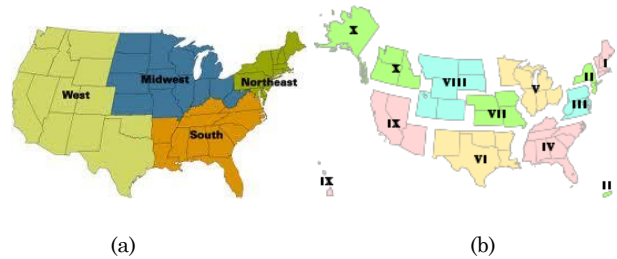

(a)  (b)

**Figure 3. Different regional breakdowns used in our region-based hierarchical classifiers. a) 4 census regions, and b) 10 standard federal regions**

## 7. HIERARCHICAL ENSEMBLE OF CLASSIFIERS

For location classification at a smaller granularity, such as city level, classifiers have to discriminate among many locations to compute the prediction. To simplify this task, a large classification problem may be divided into multiple smaller classification problems where classifiers are organized in a hierarchy. The initial classifier in such a system will compute a high-level classification, such as for time zone, and lower level classifiers will be trained for each of the classes of the high-level classifier. The low-level classifier that is used for a particular instance is determined by the classification of the initial classifier. Such a hierarchical classification scheme has been used to improve classification performance in a number of fields such as web, biology, and document analysis [Dumais et al. 2000; Sun et al. 2001].

In our work, we have developed location predictors using a two level hierarchy. We experimented with several options as the first level of hierarchy: time zone, state, and two variations of geographic regions. In all cases, city is classified at the second level.

### 7.1 Time Zone Hierarchy

When time zone is the first level of hierarchy, we classify between only the 4 US time zones (Eastern, Central, Mountain, and Pacific) since cities in our training corpus were restricted to those time zones. We first trained an ensemble time zone classifier from our training corpus using all content-based classifiers and the behavior-based classifier. City classifiers were trained for each time zone, where each classifier was limited to predicting only the cities in its time zone and trained with only examples from that time zone.

### 7.2 State Hierarchy

In this classification scheme, we use US states as the first level of the hierarchy. The ensemble state classifier contains only our content-based classifiers, city classifiers are built for all states that contain more than one city in our data set.

### 7.3 Region Hierarchy

In this classification, we use US geographical regions as the first level of hierarchy. We tried two different regional breakdowns of the US: census and federal[15] (See Figure 3 for regional breakdowns). The US Census Bureau divides the US into four regions (Northeast, Midwest, South, and West). The standard Federal Regions were established by the Office of Management and Budget and is composed of 10 regions each containing 4 to 6 adjacent states. The regional hierarchical classifiers are built using the same basic approach as for the state hierarchical classifiers.

## 8. EXPERIMENTS

We conducted many experiments to evaluate different aspects of our algorithms. Let the total number of users in our test set be n. When this is given to our location predictor, only $n_1$ predictions are correct. Hence, we define accuracy of classification as $n_1/n$.

### 8.1 Individual Classifier Performance

Table 3 shows the comparative performance of the individual location classifiers. The Place Name statistical classifier gives the best accuracy. The high accuracy of the place name-based classifier may be explained by the fact that many users send tweets containing names of places (cities and states in our system), and those place names tend to have bias towards users' home cities. The low accuracy of the visit-history classifier is due to the sparseness of the needed Foursquare URLs in our dataset (only 6.6% of tweets in our dataset contained these URLs and some of these could not resolved to a venue).

---

[15] http://en.wikipedia.org/wiki/List_of_regions_of_the_United_States

|  | Word | Hashtag | Place name | Local-place | Visit-history |
|---|---|---|---|---|---|
| Accuracy | 0.34 | 0.17 | 0.54 | 0.5 | 0.13 |

**Table 3. City-level location prediction accuracy comparison among different classifiers**

### 8.2 Ensemble Classifier Performance

We evaluate our ensemble classifier approach by comparing three alternative designs:

i) A *single statistical classifier* where words, hashtags and place names are used together as features

ii) An *ensemble of only the statistical classifiers* for words, hashtags and place names

iii) An *ensemble of the statistical and heuristic classifiers*

Table 4 shows that boosting outperformed majority voting when we constructed an ensemble of only the statistical classifiers, however the dynamically weighted ensemble performed slightly better than boosting. Table 5 shows that the dynamically weighted ensemble slightly outperformed majority voting when we constructed an ensemble of both the statistical and heuristic classifiers.

| Majority Voting | 0.51 |
|---|---|
| Boosting | 0.55 |
| Dynamically Weighted Ensemble | 0.56 |

**Table 4. City-level location prediction accuracy comparison for ensemble of statistical classifiers**

| Majority Voting | 0.55 |
|---|---|
| Dynamically Weighted Ensemble | 0.58 |

**Table 5. City-level location prediction accuracy comparison for ensemble of statistical and heuristics classifiers**

|  | Single Statistical Classifier | Ensemble of Statistical Classifiers | Ensemble of Statistical and Heuristics Classifiers |
|---|---|---|---|
| Accuracy | 0.38 | 0.56 | 0.58 |

**Table 6. Effect of ensemble for city-level location prediction**

Table 6 shows the best performance of each of our alternative designs. Observe that using an ensemble of the statistical classifiers yields higher performance than a single statistical classifier that uses the same features. This is because each feature category has its own unique discriminative ability for location classification, and in a single classifier each is weighted equally. When a separate classifier for each feature category is constructed, the different discriminating abilities of each classifier can be weighted appropriately for each instance, resulting in better classification performance. The performance of the ensemble of statistical and heuristic classifiers is superior to the other two options, suggesting that the heuristic classifiers add additional discriminative power to the ensemble. The remaining results in this paper were generated using this ensemble design.

### 8.3 Classification Performance at Multiple Location Granularities

Table 7 shows the performance of our content-based ensemble classifiers for predicting location at the level of city, time zone, state and geographic region. Performance is generally higher for classifiers that discriminate between fewer classes.

|  | City | State | Time zone | Region (federal) | Region (census) |
|---|---|---|---|---|---|
| Accuracy | 0.58 | 0.66 | 0.73 | 0.69 | 0.71 |

**Table 7. Content-based location prediction performance**

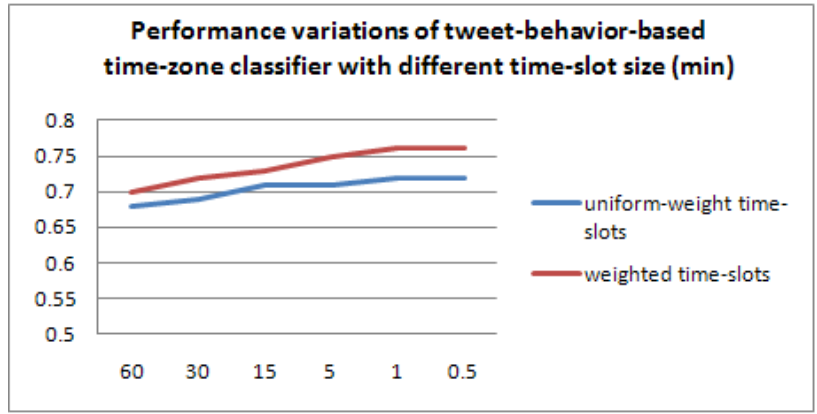

**Figure 4. Time-zone classification using tweet-behavior**

Figure 4 shows the performance of the behavior-based time zone classifier for various time slot sizes. Performance improves when time slots are weighted, and we also see an improvement in performance when the time slot size is reduced. Performance seems to level off at a slot size of 1 minute; we use that time slot size with weighting in the remainder of this paper.

|  | Content | Tweet-behavior | Combined |
|---|---|---|---|
| Accuracy | 0.73 | 0.76 | 0.78 |

**Table 8. Time-zone prediction accuracy**

Table 8 shows that we get the best time zone classification when the behavior-based classifier is combined with content-based classifiers in an ensemble (using the dynamically weighted ensemble approach).

### 8.4 Performance of Hierarchical Location Estimator

Table 9 shows the performance of different hierarchical classification approaches for city location estimation. Note that the performance of all hierarchical classifiers is superior to the single level ensemble for city prediction. The time zone based hierarchical classifier performs the best, which is largely due to the higher accuracy of predicting time zones compared to states or regions.

|  | Using time-zone hierarchy | Using State-hierarchy | Using region(federal) hierarchy | Using region (census) hierarchy |
|---|---|---|---|---|
| Accuracy | 0.64 | 0.59 | 0.6 | 0.62 |

Table 9. Performance of hierarchical city location estimator

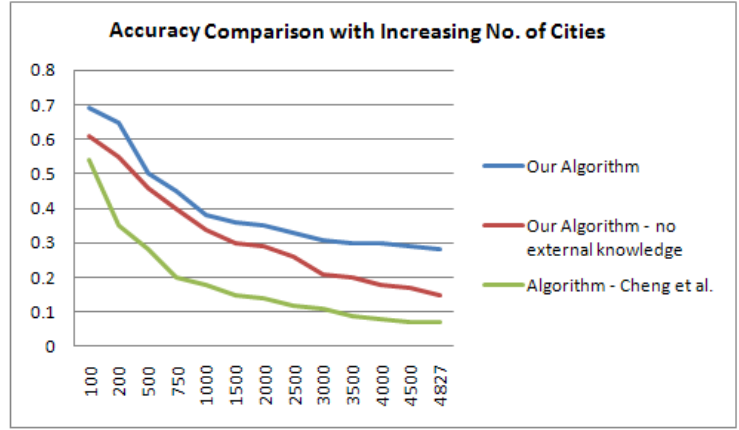

(a)

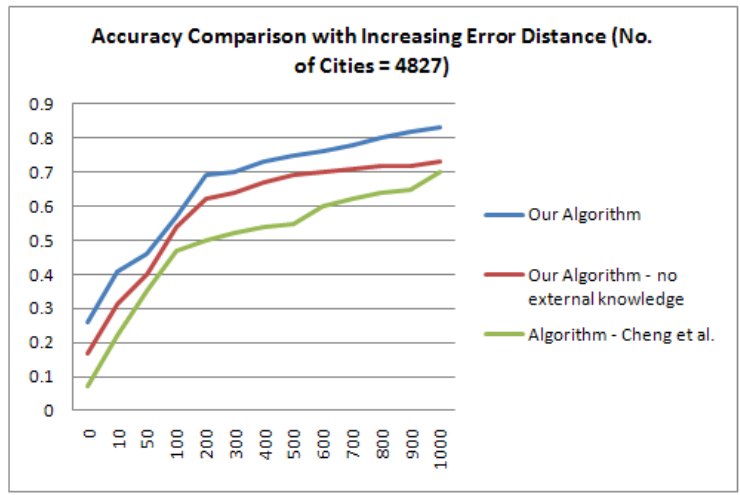

(b)

Figure 5. Comparison of our hierarchical city location predictor (with time zone hierarchy) with the best available algorithm

### 8.5 Comparison with Existing Approach

We compare the performance of our algorithm with that of Cheng et al. [2010], the previous best performing city-level location classification approach, using their dataset. We obtained Cheng et al. [2010]'s data set through correspondence with the authors.

Their dataset consists of separate training and test sets. The training set, which contains 4124960 tweets from 130689 users, consists of users who specified their location in their profile using a valid city, state pair as verified using the USGS gazetteer. The test dataset contains 5156047 tweets from 5190 users. These users reported their location in their profile as a string representing a latitude/longitude coordinate, presumably as set by their smart phones. The training dataset contains only 8 tweets containing Foursquare URLs, and the test dataset contains 10956 such tweets, representing 0.2% of the tweets in that dataset.

For comparison, we implemented their algorithm and used multiple accuracy metrics: exact accuracy and their own distance-based relaxed accuracy metric [Cheng et al. 2010]. The relaxed accuracy metric counts a location prediction as correct if it is within X miles of the actual location of the user. Figure 5 shows the performance of both algorithms: our hierarchical city location predictor with time-zone hierarchy (our best algorithm for city prediction) and our implementation of Cheng et al.'s algorithm. In particular, Figure 5(a) shows the accuracy comparison when we considered different subsets of data from the original dataset. For each subset, we first fixed N which is the number of cities (e.g. N = 500). Then, we randomly selected N cities from the list of the cities in the dataset. Then, we only included tweets from users who reported those cities as their locations in our training and test set. We tried different values of N, such as 100, 200, 500, 1000, 4827 (total no. of cities in the original dataset). We observe that, our algorithm outperforms Cheng et al.'s algorithm in all cases. Figure 5(b) shows accuracy comparison when we fixed the total number of cities N = 4827, however varied the error distance. We considered different error distances such as 0, 10, 50, 100, 1000. Our algorithm outperforms Cheng et al.'s algorithm in all cases. Since Cheng et al. did not use any external knowledge (such as a geographic gazetteer), we also compare the performance of our algorithm without the use of any external knowledge (by removing the place-name and visit-history classifiers from the ensemble). Even without external knowledge, our algorithm still has superior performance.

### 8.6 Effect of Explicit Location Reference

We were curious about the impact of the availability of explicit location references, such as place name mentions and the presence of Foursquare URLs, on classification performance. If the impact is substantial, then users can effectively mask their location by never mentioning place names. To test this, we computed the performance of just the word and hashtag statistical classifiers in an ensemble. We found that locations are still predictable, but accuracy was reduced (city level location predictor was able to predict with 0.34 accuracy without hierarchy and 0.4 accuracy with time-zone hierarchy). This suggests that users may be able to partially mask their location by being careful not to mention location names in their tweets. It may also be possible to create Twitter clients that detect location names and either warn users before posting the tweet or automatically modify the tweet to remove or obscure the location name.

### 8.7 Real World Usage Issues

There are several factors that might affect the performance of our algorithm in real world usage. First, it may not be possible to collect 200 tweets for every user, especially when tweets are collected using stream-based methods. How does accuracy change as the number of tweets for a user is decreased? We explored this scenario by capping the number of tweets for each user and found, unsurprisingly, that performance drops with the number of tweets per user. Figure 6 shows the result for city classification using a hierarchical ensemble based on time zone. Performance generally drops with decreasing tweets since with fewer tweets our classifiers lack enough features to accurately predict locations. We also compute the time required to make a location prediction, and found that our location predictor can compute the prediction for a user in less than a second (670 ms to make a prediction when 200 tweets/user is used and only 200 ms to make a prediction when 50 tweets/user is used). This is likely less than the time needed to retrieve a user's most recent tweets from Twitter, and suggests that our algorithm should be applicable in settings when reasonably accurate predictions are needed from few tweets and within a short time.

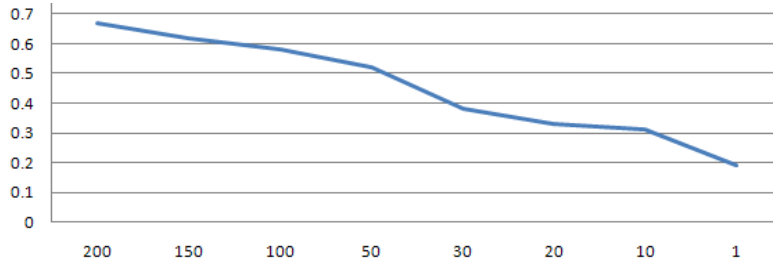

**Figure 6. Accuracy variation with decreasing total number of tweets per user**

## 9. MOVEMENT AND LOCATION PREDICTION

In our work so far, we have made an assumption that users are from the location in which we initially detected them and that they did not change locations during the period of the 200 tweets that we recorded for each user. Obviously, this assumption is unlikely to hold for all 9551 users and it is important to understand the impact of this assumption on the the results of our algorithm. In addition, if it is possible to identify the users who have travelled, then we can treat them separately and potentially improve classification performance for all users.

### 9.1 Effect of Movement on Location Prediction

To test our assumption about the location of users and understand its impact on our results, we analyzed the geographical distribution of the geo-tagged tweets in our corpus (note that geo-tags are not used in any of our prediction algorithms, although around 65% of the tweets in our dataset are geo-tagged). Table 10 and 11 show that most of our users stayed within 10 miles of the location in which we found them across all of their historical 200 tweets; location prediction accuracy is also higher for those users. This suggests that our ground truth assumption, that users are at home in the location where we identified them, is correct for most of our users.

|  | 0-10 | 11-100 | 101-500 | 500+ |
|---|---|---|---|---|
| **% of users** | 0.77 | 0.15 | 0.07 | 0.01 |
| **Accuracy** | 0.66 | 0.61 | 0.52 | 0.5 |

**Table 10. Percentage of users and prediction accuracies for different average geo-distance (miles) between tweets.**

|  | 0-10 | 11-100 | 101-500 | 500+ |
|---|---|---|---|---|
| **% of users** | 0.31 | 0.39 | 0.09 | 0.21 |
| **Accuracy** | 0.69 | 0.68 | 0.58 | 0.52 |

**Table 11. Percentage of users and prediction accuracies for different max geo-distance (miles) between tweets.**

### 9.2 Detecting Travelling Users

Based on the result on the previous subsection, it seems that a classifier for traveling users could be added to our location prediction algorithm as a pre-filtering step to eliminate traveling users. We attempt to build such a binary classifier from our data.

To train the classifier, we use the tweets with their geo-tagged information. A user was labeled as traveling if his/her maximum geo distance between tweets was above 100 miles, and not-traveling otherwise. Like our location classification approach, we used words, place names, and hashtags as features. Similar to the feature extraction method described in section 5, we tokenize all tweets in the training set to remove punctuation, white spaces, URLs and tokens containing special characters except for tokens that represent hashtags. We also apply stop word elimination and part of speech analysis using OpenNLP to identify words (marked as nouns by a part-of-speech tagger). Hashtags are all tokens starting with the # symbol, and place names are identified using USGS gazetter using the same approach described in section 5. In addition, we used time-based features that were calculated as the standard deviation of tweeting times in a particular slot of the day (24 slots for the entire day, 1 hour for each slot).

We tried with several classification algorithms from WEKA such as SMO, Naïve Bayes, Logistic Regression, J48, Random Forest. We selected SMO, which outperformed other classifiers and produced 75% F1 with 10-fold cross-validation.

### 9.3 Improving Location Prediction using Travelling Users Detection

We used the result of this classification for eliminating travelling users from our test set, which resulted in improvements in location prediction accuracy (see Table 12 and Table 13). In the future, we plan to improve the performance of our travelling user prediction algorithm, and use that to further improve the location prediction accuracies at different granularities.

|  | City | State | Time zone | Region (federal) | Region (census) |
|---|---|---|---|---|---|
| Accuracy | 0.61 | 0.70 | 0.80 | 0.72 | 0.73 |

Table 12. Location Prediction Performance when Users Classified as Travelling were Eliminated

|  | Using time-zone hierarchy | Using State-hierarchy | Using region(federal) hierarchy | Using region (census) hierarchy |
|---|---|---|---|---|
| Accuracy | 0.68 | 0.62 | 0.63 | 0.64 |

Table 13. Hierarchical Location Prediction Performance when Users Classified as Travelling were Eliminated

### 10. CONCLUSION

In this paper, we have presented a hierarchical ensemble algorithm for predicting the home location of Twitter users at different granularities. Our algorithm uses a variety of different features, leverages domain knowledge and combines statistical and heuristics classifications. Experimental performance demonstrates that our algorithm achieves higher performance than any previous algorithms for predicting locations of Twitter users. We identify several avenues of future research. First, we are interested to apply our method to predict location at even smaller granularities, such as the neighborhood level. Towards that, we plan to incorporate more domain knowledge in our location prediction models, such as a landmark database[16]. Along the same line, it would be interesting to explore the possibilities of predicting the location of each message. Second, we plan to investigate further on detecting travelling users, and use that to improve the accuracies of our location classifiers. Third, we would like to support incremental update of our models for better integration with streaming analytics applications. Finally, we hope to integrate our algorithm into various applications to explore its usefulness in real world deployments.

---

[16] http://poidirectory.com/poifiles/united_states/